# SEMANTIC RECONSTRUCTION OF CONTINUOUS LANGUAGE FROM MEG SIGNALS

*Bo Wang[1], Xiran Xu[1], Longxiang Zhang[1], Boda Xiao[2], Xihong Wu[1,3], Jing Chen[1,3]*

[1]Key Laboratory of Machine Perception (Ministry of Education), Speech and Hearing Research Center,
School of Intelligence Science and Technology, Peking University
[2]Academy for Advanced Interdisciplinary Studies, Peking University
[3]National Biomedical Imaging Center, College of Future Technology, Peking University
*janechenjing@pku.edu.cn*

## ABSTRACT

Decoding language from neural signals holds considerable theoretical and practical importance. Previous research has indicated the feasibility of decoding text or speech from invasive neural signals. However, when using non-invasive neural signals, significant challenges are encountered due to their low quality. In this study, we proposed a data-driven approach for decoding semantic of language from Magnetoencephalography (MEG) signals recorded while subjects were listening to continuous speech. First, a multi-subject decoding model was trained using contrastive learning to reconstruct continuous word embeddings from MEG data. Subsequently, a beam search algorithm was adopted to generate text sequences based on the reconstructed word embeddings. Given a candidate sentence in the beam, a language model was used to predict the subsequent words. The word embeddings of the subsequent words were correlated with the reconstructed word embedding. These correlations were then used as a measure of the probability for the next word. The results showed that the proposed continuous word embedding model can effectively leverage both subject-specific and subject-shared information. Additionally, the decoded text exhibited significant similarity to the target text, with an average BERTScore of 0.816, a score comparable to that in the previous fMRI study.

## 1. INTRODUCTION

Every year, a considerable number of people lose their ability to speak due to cerebral stroke or ALS (Amyotrophic Lateral Sclerosis). Over the past few decades, brain-computer interfaces (BCIs) have made great progress in language decoding. Previous studies have demonstrated that acoustic information [1], [2], articulatory movements [3], [4], and semantic information [5] could be effectively decoded from intracranial recordings, offering hope for restoring communication to the patients. However, since invasive, these BCIs were not suitable for the majority of the patients.

Decoding semantic of language from non-invasive recordings, on the other hand, remains a major challenge. Research utilizing functional magnetic resonance imaging (fMRI) has demonstrated the robust decoding of semantic information from the blood-oxygen-level dependent (BOLD) response [6], [7]. Nonetheless, fMRI lacks portability and is unable to capture rapid changes, making it impractical for real-time applications in daily life [8]. Instead, Electro-/Magnetoencephalography (EEG/MEG) measures neural activity at millisecond resolution with a safe and potentially wearable setup [9], making them more suitable for BCI application.

However, decoding semantic from EEG/MEG is extremely challenging due to its low signal-to-noise ratio (SNR) and low spatial resolution. Previous studies typically used a subject-specific or common linear decoder to regress semantic features from EEG/MEG data evoked by a single word [10]–[14]. Subsequently, the most probable word from a small closed set was selected as the decoding output based on the distance between the reconstructed features and the true features. In work of Hultén et al. [11], MEG data were recorded when subjects were reading the written text of 118 nouns. Subject-specific ridge linear models were adopted to predict word2vec features from MEG data using leave-two-out pairwise comparisons (the model was trained on 116 of the words and tested on the remaining 2). A maximum mean pair-wise decoding accuracy of 66% (chance accuracy: 50%) was reported with a decoding window of 100 ms. In a more recent work of Ghazaryan et al. [12], the similar paradigm was used but with a smaller word set (60 words), and the decoding accuracy significantly higher than chance level was only shown on 6 of 19 subjects. When averaging MEG data across subjects to improve signal SNR, the decoding accuracy was up to 78%.

The limited amount of data and simple linear models impose constraints on the effective modeling of the intricate relationship between neural responses and features. Moreover, since previous studies showed that neural representation of semantic shared certain commonalities but also displayed a diversity across individuals [15], [16], the common or subject-specific decoder was unable to capture both shared and subject-specific neural response patterns to the same stimulus. As a result, the decoding performance was limited, and it was only possible to select words from a closed set, rather than open vocabulary decoding.

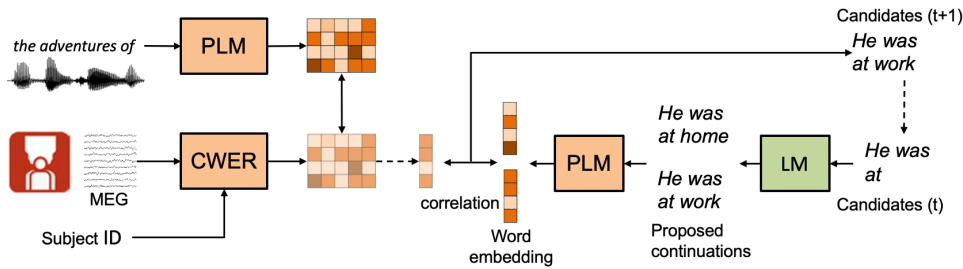

**Fig. 1** Architecture of generating word sequences from MEG with a CWER model and beam search decoding.

To address these issues, we proposed a novel framework to decode semantic from MEG signals which were recorded when the subjects were listening to speech, i.e., spoken stories. Our framework consists of two parts, as shown in Fig. 1. Firstly, a continuous word embedding reconstruction (CWER) model is trained to reconstruct continuous word embeddings from MEG. The continuous word embeddings were derived from temporal sequence of spoken words and transformer-based pre-trained language model (PLM) which encodes important linguistic information, including semantic features, syntactic features, and long-distant dependencies [17], [18]. Neuroimaging studies have indicated that these word embeddings could be effectively mapped to the neural activity measured during natural speech processing [5], [19], [20]. Subsequently, the similarity was measured between the reconstructed word embeddings and the word embeddings of the subsequent words which are predicted by a language model (LM) for a candidate word sequence. The similarity was then used as a measure of the probability of the next word. By incorporating this probability into a beam search algorithm, we can generate the most probable sequence of words.

Compared with the previous studies [10]–[14], our framework models the relationship between neural responses and semantic features in a data-driven manner through neural networks. By adding a subject embedding layer, the reconstruction model captures both the subject-specific and the common patterns of the neural responses to the same piece of speech. In terms of the decoding output, instead of selecting a word from a small closed set, our framework could generate open vocabulary continuous word sequence.

## 2. METHODS

### 2.1. Continuous word embedding reconstruction model

The CWER model aimed to reconstruct word embeddings from a sequence of high-dimensional MEG signals. These MEG signals were recorded when healthy subjects were listening passively to stories in their native language. The story text was fed into PLM, and the hidden layers output was utilized as the target of word embeddings. With the temporal boundaries of words, continuous word embeddings can be created by filling word embeddings in their corresponding period.

Consider $X \in R^{C \times T}$ as a segment of MEG data from subject $s \in [S]$, where $C$ represents the number of MEG sensors, $T$ denotes the number of time samples, and $S$ is the subject number. Let $Z \in R^{D \times T}$ be a segment of continuous word embeddings of the story text that was corresponding to the MEG segment. Here, $D$ represents the dimensionality of the word embeddings, and both $X$ and $Z$ share the same time samples. The CWER model takes $X$ and $s$ as input and outputs the estimated $\hat{Z}$. Previous study revealed that MEG response to spoken words showed dynamic spatial patterns [21]. As 1D convolution can effectively capture temporal dependencies while extracting spatial patterns among channels, it was used in the present study. An overview diagram of the CWER model is shown in Fig. 2. The MEG data $X$ is first fed in a linear layer and followed by a $1 \times 1$ convolution layer to transform MEG signals into a higher hidden space with the dimension of $D_1$. Then, a subject embedding layer is added, conditioning on the subject's one-hot label, to learn a linear transformation $M_s \in R^{D_1 \times D_1}$ along the channel dimension for each subject. This allows us to account for inter-subject variation. Subsequently, a stack of five blocks of three convolutional layers was applied to extract MEG spatial-temporal features. The convolutional layer was the same as that in previous work [22] but with a kernel size of 9 over the time axis to further increase the receptive field. Finally, the output of convolutional blocks was sequentially fed into a $1 \times 1$ convolution layer with $2D_2$ channels, followed by a GELU activation, and then another $1 \times 1$ convolution layer with $D$ channels to output the estimated $\hat{Z}$.

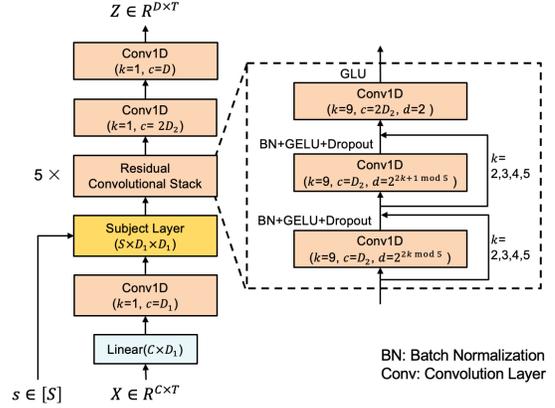

**Fig. 2** Architecture of the CWER model.

The reconstructed embeddings were expected to be similar to the target embeddings while being dissimilar to the non-target embeddings, enabling effective selection of target words from the set during the subsequent beam search decoding. Therefore, a contrastive loss was adopted. More specifically, for a segment of MEG data $X$, the corresponding word embedding $Z$ is considered as a positive sample (denoted as $Z^1$), and $N - 1$ non-target word embeddings segments $\{Z^2, \ldots, Z^N\}$ sampled from the training set are the negative samples. Here, $Z^i$ ($i = 1, 2, \ldots, N$) comprises a series of $T$ $D$-dimensional vectors, i.e., $Z^i = (z_1^i, z_2^i, \ldots, z_T^i)$. The InfoNCE loss is considered in this paper:

$$L = -\sum_{t=1}^{T} \log \frac{\exp(\text{sim}(\hat{z}_t, z_t^1)/\tau)}{\sum_{i=1}^{N} \exp(\text{sim}(\hat{z}_t, z_t^i)/\tau)} \quad (1)$$

where $\text{sim}(\cdot, \cdot)$ is Pearon correlation between two vectors, and $\tau$ is the temperature parameter in InfoNCE [23]. Intuitively, this loss is the log loss of an N-way softmax-based classifier that tries to classify $\hat{Z}$ as $Z^1$.

### 2.2. Word sequence generation

The most likely word sequence could be estimated using a beam search algorithm [6]. Specifically, the beam contains $k$ candidates. For each candidate, the language model uses the last 8-s generated words $(s_{n-i}, \ldots, s_{n-1})$ to predict the next word distribution $P(s_n|s_{n-i}, \ldots, s_{n-1})$ over the decoder vocabulary. Nucleus sampling is used to identify subsequent words that belong to the top $p$ percent of the probability mass and have a probability within a factor $r$ of the most likely word. For each continuation $s_n$, a sequence $(s_{n-L-1}, \ldots, s_n)$ consisting of the word itself and its preceding $L$-1 words is fed into the PLM model, to get the word embedding for $s_n$. Given the neural response $X$, the CWER model outputs an estimated embedding $\hat{Z}$. With the known onset time $t_{on}$ and offset time $t_{off}$ of the next word, the reconstructed word embedding for the continuations is given as:

$$\hat{z} = \frac{1}{t_{off} - t_{on}} \sum_{t=t_{on}}^{t_{off}} \hat{z}_t \quad (2)$$

As a result, the Pearson correlation between $\hat{z}$ and embeddings of all continuations are measured as the next word probability. The $k$ most likely continuations across all candidates are retained in the beam. After iterating through all the words along time, the candidate with the highest probability was output as the decoded sequence.

## 3. EXPERIMENTS

### 3.1 Dataset

The dataset used is from Donders Institute for Brain [24]. The MEG data were recorded with a 275-channel axial gradiometer CTF system at a sampling of 1200 Hz from 3 participants, while they listened to audiobooks of *The Adventures of Sherlock Holmes* in English. For each participant, the data were recorded in 10 separate sessions, consisting of 66 trials in total, with a total duration of approximately 10 hours.

### 3.2. MEG preprocessing and continuous word embedding preparation

The MEG data underwent average re-referencing and were subjected to band-pass filtering (1-40 Hz), notch filtering (49-51, 99-101, and 149-151 Hz), and down-sampling to 120 Hz. Subsequently, independent component analysis (ICA) was applied for each trial to remove eye-blink artifacts. In our pilot study, we observed that MEG exhibited the highest efficiency in predicting word embedding within the low-frequency range. The MEG data were further low-pass filtered at a cut-off frequency of 4 Hz, downsampled to 40 Hz to reduce computational costs, and standardized along the time axis for each sensor.

To get continuous word embedding, each word along with its preceding $L$-1 words is fed into GPT [25]. The hidden output of the 9th layer was used as word embedding. The GPT was initialized with the model weight from the model that was used in [6] and fine-tuned on another 4 books (*The Case-Book of Sherlock Holmes, The Memoirs of Sherlock Holmes, The Return of Sherlock Holmes* and *His Last Bow*) from *Conan Doyle*. The temporal onset and offset of each spoken word were included in the dataset. To create continuous word embedding, a multivariate time series was constructed with each word embedding filling in its corresponding time slot. To match the MEG data, the continuous word embedding was at a sampling rate of 40 Hz and 4 Hz low-pass filtered.

### 3.3 Experiment setup

For each subject, the MEG data from session 4 (8 trials) were used as the testing set, and the data from the remaining 9 sessions (58 trials) were used as the training set. To compensate the delay between stimulus and its corresponding brain response, the MEG data were shifted backward by 0.25 s firstly. The MEG data and the paired continuous word embeddings in the training set were further split into segments with a duration of 10 seconds and an overlap of 80%. The data in the testing set were split into segments with a duration of 10 seconds without overlapping.

During the training of the CWER model, an Adam optimizer with a learning rate of $5 \times 10^{-5}$ was used. The batch size was set to 32, and the negative sample $N$ was set to 128. The hidden size for both $D_1$ and $D_2$ was set to 256. The dropout rate for residual convolutional layers was set to 0.5. The temperature parameter in InfoNCE was set to 0.025. Training was stopped when no loss reduction was found for 2 consecutive training epochs in the testing set. For comparison, a CWER model without subject layer and three subject-specific CWER models for every subject were also trained.

Besides, a linear model was used as the baseline. The linear mapping was trained with ridge regression to reconstruct word embedding $z_t$ from the time series of MEG data $\{x_{t+\tau_1}, ..., x_t ..., x_{t+\tau_2}\}$. Here, $\tau_1$ and $\tau_2$ were set to $-40$ and $60$, corresponding to 1-s before and 1.5-s after stimuli respectively, as a previous study suggested that brain signal robustly encoded upcoming words starting $-1$-s before and ending 1.5-s after the onset of spoken words [5]. The ridge parameter was determined by cross-validation procedure with a grid search (20 grid values were logarithmically spaced ranging from $10^{-3}$ to $10^5$). More details about the linear model can be found in [26]. For comparison, a common linear model and three subject-specific linear models were trained, respectively. All the models are implemented with Pytorch.

During the beam search decoding, the fine-tuned GPT in *Continuous word embedding preparing* was served as the language model and the word embedding extraction model. The beam width was set to 200, and the nucleus sampling parameters $p$ and $r$ were set to 0.9 and 0.1, respectively. The words that occurred at least twice in the train set were selected, forming a decoder vocabulary consisting of 3,660 unique words. The constraint on the continuation number for each candidate is the same as that in [6].

### 3.4 Evaluation

*3.4.1 Segment-level evaluation*

The evaluation at the segment level was treated as an assessment of a retrieval task. With the trained CWER model, the continuous word embedding was reconstructed for each trial in the testing set. These reconstructed continuous word embeddings were then split into segments with a duration of 3, 5, and 10 s without overlapping, resulting in 1,210, 723, and 359 segments for each duration condition, respectively. The same segmentation manipulation was also applied to the true continuous word embeddings, thereby creating a set of candidates.

For each reconstructed segment, the Pearson correlations between the reconstructed segment and each segment in the candidate set were calculated, and then the top 10 segments in the candidate set with the highest correlation were retrieved. Top-10 accuracy was defined as the percentage of reconstructed segments whose target segment was in the retrieval set. Meanwhile, rank accuracy was also calculated as the previous study did [7]. For each reconstructed segment, the rank of its target segment among the candidate set was normalized into rank accuracy using equation (3).

$$\text{rank}_{acc} = 1 - \frac{<\text{rank}> - 1}{ - 1} \quad (3)$$

*3.4.2 Sequence-level evaluation*

To assess the quality of the generated text, BERTScore [27] was used to evaluate semantic similarity between the generated text and the target text. Window similarity was measured by scoring the predicted and target words with BERTScore within a 20-s window around every second of a trial. Trial similarity was then calculated by averaging BERTScore across all windows in a trial.

To get the chance level of decoding score, null sequences were generated by using only language model. During beam search decoding, the next word probability was randomly assigned instead of being estimated from MEG data. For each trial in the testing set, 500 null sequences were generated. The $p$-value was given by the proportion of null distribution scores that were higher than the brain decoded. The decoding accuracy for trial (window) was defined as the percentage of trial (window) whose $p$-value was smaller than 0.05.

## 4. RESULT AND DISCUSSION

### 4.1 Continuous word embedding reconstruction

Table 1. Decoding accuracy for segment level evaluation of each model. (s. s.: subject specific; w.o. s.l.: without subject layer)

| Model | Top-10 acc (%) | | | Rank acc (%) | | |
|---|---|---|---|---|---|---|
| | Duration (s) | | | Duration (s) | | |
| | 3 | 5 | 10 | 3 | 5 | 10 |
| Random | 0.1 | 1.3 | 2.7 | 50.1 | 50.1 | 50.2 |
| Linear (common) | 10.7 | 25.8 | 64.8 | 82.5 | 88.4 | 95.3 |
| Linear (s. s.) | 17.2 | 39.2 | 83.2 | 87.0 | 92.4 | 97.8 |
| CWER (w.o. s. l.) | 24.4 | 49.7 | 85.4 | 85.0 | 93.0 | 98.1 |
| CWER (s. s.) | 26.3 | 50.2 | 85.6 | **90.7** | 94.5 | **98.3** |
| CWER | **30.0** | **55.7** | **89.0** | 90.3 | **94.7** | **98.3** |

The segment-level accuracy for each model as a function of the decoding duration is shown in Table 1. The accuracy for the random model was calculated by using a random continuous word embedding and the results were averaged over 500 runs. As expected, the accuracies of the common linear model were higher than that of the random model, indicating that the semantic features could be reconstructed from MEG with a linear decoder. The subject-specific linear model outperformed the common linear model, suggesting significant individual differences among the subjects. Our common CWER model (without subject layer) and subject-specific CWER model outperformed the common linear model and subject-specific linear model, respectively. This showed that the utilization of a non-linear network can enhance decoding accuracy. Moreover, the CWER model outperformed both the common CWER model and the subject-specific CWER model, which revealed that the CWER model can effectively leverage both subject-specific and subject-shared information.

Compared to the previous studies of semantic decoding with MEG or fMRI [7], [10]-[14], [28], the performance of our proposed models showed the promising advantage on the segment-level evaluation. The decoding performance improvement can be attributed to the fact that large amount of the training data and nonlinear CWER model were used in the current work.

To determine whether words can be distinguished as concrete or abstract from the reconstructed word embeddings, 24 concrete words and 25 abstract words from the testing set were selected. The target and reconstructed word embeddings were decomposed via a 2-dimensional tSNE [29]. The result is shown in Fig. 3. As expected, the concrete words and abstract words can be well distinguished in the reconstructed word embeddings tSNE space, confirming the good performance on semantic reconstruction from MEG signals.

### 4.2 Word sequence generation

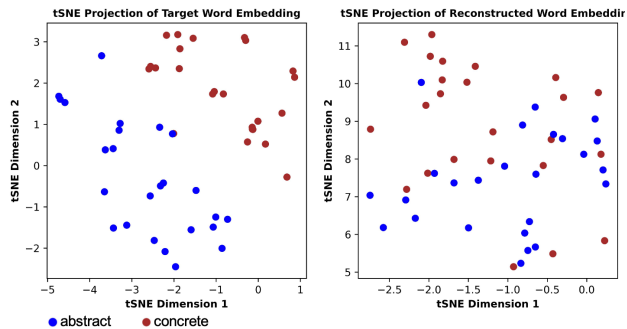

**Fig. 3** TSNE plot visualizing of abstract and concrete words for the target and the reconstructed word embedding.

Table 2. Similarity scores between the generated word sequence and the target word sequence, and the decoding accuracy for windows and trials. The scores and accuracies were averaged across all windows or trials of all subjects.

| Model | Score | Acc (%) | |
|---|---|---|---|
| | | Window | Trial |
| Null | 0.812 | – | |
| CWER (w.o. s. l.) | 0.815 | 18.1 | 79.8 |
| CWER | **0.816** | **20.8** | **100** |

Metrics scores for sequence-level evaluation of each model are shown in Table 2. Scores for Null were computed by averaging metrics for 500 null sequences. For the CWER model, the averaged BERTScore was 0.816, that was higher than the score for the null model and was also higher than that in the previous fMIR work where a BERTScore of 0.810 was reported [6]. The similarity between the generated and target text was significant for all trials, indicating effective semantic reconstruction from the brain. Additionally, the CWER model outperformed the CWER model without a subject layer. This revealed that improving continuous word embedding reconstruction performance can help generate text with better contextual similarity.

The decoding accuracy was 20.8% for windows, which indicated that 20.8% windows had a significant BERTScore. The value was lower than that in [6], where 72-82% timepoints had a significantly higher BERTScore than chance level. This is because the semantic of language is spatially distributed in the cerebral cortex [16]. The previous study utilized fMRI data, which has a significantly higher spatial resolution compared to the MEG data used in this study.

It should be noted that the actual timing information of words was used during sequence generation. However, in BCI applications, the timing information of words is typically unknown. In such instances, a model can be developed as previous studies to predict the onset of words from MEG data, as those methods used in the previous studies [4], [6]. Simultaneously, given the significant advantage of MEG in temporal resolution compared to fMRI, it is more effective in extracting highly dynamic acoustic features or articulatory movements. This information would assist in generating text from the brain. It is worthy to be investigated in future work.

### 5. CONCLUSION

In the present work, we proposed a framework to reconstruct semantic of language from MEG data. The semantic feature reconstruction performance was greatly improved by using a neural network that leveraged both subject-specific and subject-shared information. By using a beam search language model, text sequences were generated from the reconstructed features. The generated sequences demonstrated significant similarity with the target sequences. The result of this work could support the semantic decoding from non-invasive MEG.

### 6. ACKNOWLEDGEMENTS

This work was supported by the National Key Research and Development Program of China (No.2021ZD0201503), a National Natural Science Foundation of China (No.12074012), and the High-performance Computing Platform of Peking University.

## 7. REFERENCES

[1] H. Akbari, B. Khalighinejad, J. L. Herrero, A. D. Mehta, and N. Mesgarani, "Towards reconstructing intelligible speech from the human auditory cortex," *Sci Rep*, vol. 9, no. 1, p. 874, 2019.

[2] D. A. Moses, N. Mesgarani, M. K. Leonard, and E. F. Chang, "Neural speech recognition: continuous phoneme decoding using spatiotemporal representations of human cortical activity," *J. Neural Eng.*, vol. 13, no. 5, p. 056004, 2016.

[3] F. R. Willett et al., "A high-performance speech neuroprosthesis," *Nature*, vol. 620, no. 7976, pp. 1031–1036, 2023.

[4] D. A. Moses et al., "Neuroprosthesis for decoding speech in a paralyzed person with anarthria," *New England Journal of Medicine*, vol. 385, no. 3, pp. 217–227, 2021.

[5] A. Goldstein et al., "Shared computational principles for language processing in humans and deep language models," *Nat Neurosci*, vol. 25, no. 3, pp. 369–380, 2022.

[6] J. Tang, A. LeBel, S. Jain, and A. G. Huth, "Semantic reconstruction of continuous language from non-invasive brain recordings," *Nat Neurosci*, vol. 26, no. 5, pp. 858–866, 2023.

[7] F. Pereira et al., "Toward a universal decoder of linguistic meaning from brain activation," *Nat Commun*, vol. 9, no. 1, p. 963, 2018.

[8] N. K. Logothetis, "What we can do and what we cannot do with fMRI," *Nature*, vol. 453, no. 7197, pp. 869-878, 2008.

[9] R. A. Seymour et al., "Using OPMs to measure neural activity in standing, mobile participants," *NeuroImage*, vol. 244, p. 118604, 2021.

[10] M. Toneva, O. Stretcu, B. Poczos, L. Wehbe, and T. M. Mitchell, "Modeling task effects on meaning representation in the brain via zero-shot MEG prediction," in *Proceedings of the 34th International Conference on Neural Information Processing Systems (NeurIPS 2020)*, pp. 5284–5295, 2020.

[11] A. Hultén et al., "The neural representation of abstract words may arise through grounding word meaning in language itself," *Human Brain Mapping*, vol. 42, no. 15, pp. 4973-4984, 2021.

[12] G. Ghazaryan et al., "Trials and tribulations when attempting to decode semantic representations from MEG responses to written text," *Language, Cognition and Neuroscience*, 2023.

[13] S. Alizadeh, H. Jamalabadi, M. Schönauer, C. Leibold, and S. Gais, "Decoding cognitive concepts from neuroimaging data using multivariate pattern analysis," *NeuroImage*, vol. 159, pp. 449–458, 2017.

[14] A. M. Chan, E. Halgren, K. Marinkovic, and S. S. Cash, "Decoding word and category-specific spatiotemporal representations from MEG and EEG," *NeuroImage*, vol. 54, no. 4, pp. 3028-3039, 2011.

[15] S. L. Frisby, A. D. Halai, C. R. Cox, M. A. L. Ralph, and T. T. Rogers, "Decoding semantic representations in mind and brain," *Trends in Cognitive Sciences*, vol. 27, no. 3, pp. 258–281, 2023.

[16] A. G. Huth, W. A. de Heer, T. L. Griffiths, F. E. Theunissen, and J. L. Gallant, "Natural speech reveals the semantic maps that tile human cerebral cortex," *Nature*, vol. 532, no. 7600, pp. 453-458, 2016.

[17] G. Jawahar, B. Sagot, and D. Seddah, "What does BERT learn about the structure of language?," in *Proceedings of the 57th Annual Meeting of the Association for Computational Linguistics (ACL 2019)*, pp. 3651–3657, 2019.

[18] I. Tenney et al., "What do you learn from context? Probing for sentence structure in contextualized word representations," In: *Proceedings of the 7th International Conference on Learning Representations (ICLR 2019)*, 2019.

[19] C. Caucheteux and J.-R. King, "Brains and algorithms partially converge in natural language processing," *Commun Biol*, vol. 5, no. 1, pp. 1-10, 2022.

[20] M. Toneva and L. Wehbe, "Interpreting and improving natural-language processing (in machines) with natural language-processing (in the brain)," in *Proceedings of the 33rd International Conference on Neural Information Processing Systems (NeurIPS 2019)*, pp. 14954-14964, 2019.

[21] F. Pulvermüller, Y. Shtyrov, and R. Ilmoniemi, "Spatiotemporal dynamics of neural language processing: an MEG study using minimum-norm current estimates," *NeuroImage*, vol. 20, no. 2, pp. 1020–1025, 2003.

[22] A. Défossez, C. Caucheteux, J. Rapin, O. Kabeli, and J.-R. King, "Decoding speech from non-invasive brain recordings," *arXiv preprint arXiv:2208.12266*, 2022.

[23] O. Henaff, "Data-efficient image recognition with contrastive predictive coding," in *Proceedings of the 37th International Conference on Machine Learning (ICML 2020)*, pp. 4182–4192, 2020.

[24] K. Armeni, U. Güçlü, M. van Gerven, and J.-M. Schoffelen, "A 10-hour within-participant magnetoencephalography narrative dataset to test models of language comprehension," *Sci Data*, vol. 9, no. 1, p. 278, 2022.

[25] A. Radford, K. Narasimhan, T. Salimans, and I. Sutskever, "Improving language understanding by generative pre-training," 2018.

[26] M. J. Crosse, G. M. Di Liberto, A. Bednar, and E. C. Lalor, "The multivariate temporal response function (mTRF) toolbox: A MATLAB toolbox for relating neural signals to continuous stimuli," *Front. Hum. Neurosci.*, vol. 10, art. 604, 2016.

[27] T. Zhang, V. Kishore, F. Wu, K. Q. Weinberger, and Y. Artzi, "BERTScore: Evaluating Text Generation with BERT," In: *Proceedings of the 8th International Conference on Learning Representations (ICLR 2020)*, 2020.

[28] J. Wang, V. L. Cherkassky, and M. A. Just, "Predicting the brain activation pattern associated with the propositional content of a sentence: Modeling neural representations of events and states," *Human Brain Mapping*, vol. 38, no. 10, pp. 4865–4881, 2017.

[29] L. van der Maaten and G. Hinton, "Visualizing data using t-SNE," *Journal of Machine Learning Research*, vol. 9, no. 86, pp. 2579–2605, 2008.